\documentstyle[aps,epsfig]{revtex}                         

\begin{document}

\twocolumn

\title{Comment on ''Matter-Wave Interferometer for Large Molecules''
       }

\maketitle

In a recent Letter \cite{brezgen02} Brezgen et al. claim that a
classical model of interactions is in principle unsuitable to
reproduce interferometer experiments with C$_{70}$ molecules, and
that an optical model based on the de Broglie waves of the
molecule explains the experimental data. In their Letter they
clarify to a certain extent the objections raised by one of the
authors \cite{hofer01} concerning their previous publication
\cite{arndt99}. The clarification is much appreciated. However, it
is not fully convincing.

To support their claim Brezgen et al. compute the visibility of
the interference fringes based on two separate models: (i) A
'classical' model from geometrical optics, in which the movement of the molecules
through a succession of optical gratings is described as rigid rays casting a
shadow pattern at the detector plane;
and (ii) a 'quantum' model, where the scalar potential of the grating
leads to a phase shift of the de Broglie wave associated with the
molecule. Additionally, the consequences of including a van der Waals term which gives a shift in scalar potentials is analyzed within both models.

\begin{figure}
\epsfxsize=0.95\hsize \epsfbox{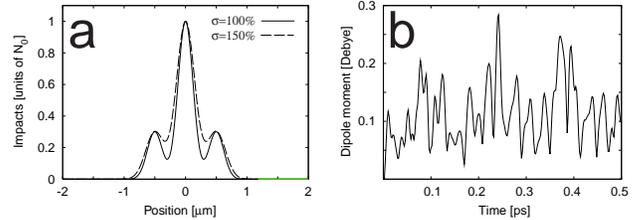} \vspace{0.5 cm}
\caption{(a) Simulation of the impacts of C$_{60}$ molecules in a
three slit system for narrow (solid line) and wide (dashed line)
effective slit-widths. This figure exactly reproduces the
intensity measurements in Ref. [3]. (b) Dipole moment of C$_{60}$
molecules at 900 K calculated from first-principle density
functional theory simulations.
        }
\label{fig001}
\end{figure}

These models are simplistic for various reasons. Most importantly,
the main conclusion that the visibility of the interference
fringes depends on the quantum character of the description is
only based on the observation that the classical model does not
have any term depending on the velocity of the molecules. But this
is a deficiency of the model rather than something characteristic
for the movement of molecules on classical trajectories. When the
temperature of a molecular beam is increased, its changed velocity
and translational energy will change its spatial distribution.
This statistical effect has been analyzed in \cite{hofer01}. Fig.
\ref{fig001} (a) shows that it results in a velocity dependence of
the visibility of the interference fringes of the earlier
experiments in \cite{arndt99} within an entirely classical model
as far as the movement of the atomic nuclei is concerned. This is
no surprise for quantum chemists or theoretical solid state
physicists, who are routinely using the Born-Oppenheimer
approximation. In surface science or biochemistry such
calculations have been successfully applied over the last decades
to describe atomic movement with classical Lagrangians in the
field of their electrons.

The description of van der Waals forces is problematic as well. In
atom interferometry van der Waals forces are known to change the
effective slit widths of diffraction gratings. But these
descriptions can only be used as a good approximation for the long
range forces, when there are no larger interaction terms. At high
temperatures molecular vibrations lead to a fluctuating molecular
dipole moment. We have computed the dipole moment of C$_{60}$
molecules  at the temperature of the experiments in \cite{arndt99}
by first principles density functional theory (see Fig.
\ref{fig001} (b))\cite{vasp}. As can be seen, the molecular dipole
moment oscillates periodically along its path. These fluctuations
are much larger in their amplitudes and time periods than the also
time-dependent polarizations of the electronic shells of atoms
responsible for van der Waals interactions. Whether the
interactions of these fluctuating molecular dipoles with the
diffraction gratings have a significant effect on the visibility
of the interference fringes can only be answered, when a detailed
model of the electronic structure of the surface of these gratings
is provided.

\begin{figure}
\epsfxsize=0.95\hsize \epsfbox{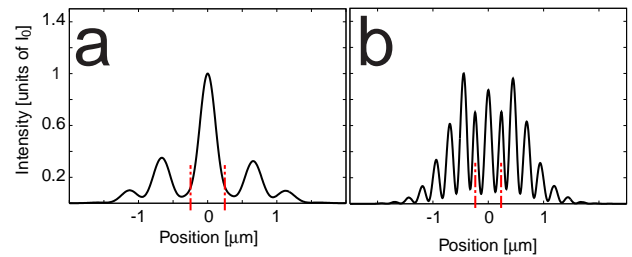} \vspace{0.5 cm}
\caption{(a) Simulated interference pattern of a photon
interferometer if the wavelength is equal to the distance between
slits. (b) Simulated interference pattern if the wavelength is
half the distance.
        }
\label{fig002}
\end{figure}

While Brezgen et al publish their visibility data, they do {\em
not} publish another crucial information: the local distribution
of maxima and minima with varying velocity. The omission does not
help to make the paper convincing, because this is a decisive
information for the discrimination between scattering and
interference. If the velocity of the molecule is increased by
100\%, then the number of visible interference peaks, given a
constant grating, should be double the original number (see Fig.
\ref{fig002}). It is difficult to obtain such a result, if the
molecules are merely scattered.

We conclude that in order to really elucidate the physics behind
these interference experiments with large molecules, an atomistic
description of the molecular trajectories is needed, including all
the interaction and excitation effects on their paths through the
system, rather than just putting labels like 'classical' or
'quantum' on models based only on geometrical projections and wave
propagation. We also conclude that crucial information is missing
in the account of the experiments.

\vspace{0.5 cm}

\noindent
R. Stadler$ ^{1}$ and W.A. Hofer$^{1}$\\
\indent $^{1}$Dept. of Physics and Astronomy \\
\indent University College London \\
\indent Gower Street, WC1E 6BT, UK

%

\end{document}